\makeatletter
\newcommand{\reserveinserts}[1]{\relax}
\makeatother

\documentclass[AMA,Times2COL]{WileyNJDv5} 

\usepackage{afterpage}

\articletype{Research Article}%

\received{Date Month Year}
\revised{Date Month Year}

\accepted{Date Month Year}
\journal{NMR in Biomedicine}
\volume{00}
\copyyear{2026}
\startpage{1}
\historydates{}
\doiheadtext{}
\usepackage{bm}

\renewcommand{\vec}[1]{\mathbf{#1}}
\newcommand{\op}[1]{\mathbf{#1}}

\newcommand{\trans}{{\scriptstyle \boldsymbol{\mathsf{T}}}}

\def\CXX{{C\nolinebreak[4]\hspace{-.05em}\raisebox{.4ex}{\tiny\bf ++}}}
\usepackage{setspace}
\usepackage[product-units=single,list-units=single,separate-uncertainty = true]{siunitx}

\newcommand{\rev}[1]{#1}

\begin{document}

\title{\rev{MRpro: open framework for model-based, learned, and quantitative MR imaging -- application to low-field MRI}}

\author[1]{Felix Frederik Zimmermann*}

\author[1]{Patrick Schuenke*}
\author[1,3]{Christoph S. Aigner}
\author[1]{Bill A. Bernhardt}
\author[1]{Mara Guastini}
\author[1]{Johannes Hammacher}
\author[4]{Helge Herthum}
\author[2]{Noah Jaitner}
\author[1]{Andreas Kofler}
\author[1]{Leonid Lunin}
\author[1]{Stefan Martin}
\author[1]{Catarina Redshaw Kranich}
\author[2]{Jakob Schattenfroh}
\author[1]{David Schote}
\author[2]{Yanglei Wu}
\author[1]{Christoph Kolbitsch}

\authormark{Zimmermann \textsc{et al}}

\address[1]{\orgdiv{}\orgname{Physikalisch-Technische Bundesanstalt (PTB)}, \orgaddress{Braunschweig and Berlin, \country{Germany}}}

\address[2]{\orgdiv{Department of Radiology}, \orgname{Charité -- Universitätsmedizin Berlin, Corporate Member of Freie Universität Berlin, Humboldt-Universität zu Berlin, and Berlin Institute of Health}, \orgaddress{\city{Berlin}, \country{Germany}}}

\address[3]{\orgdiv{Max Planck Research Group MR Physics}, \orgname{Max Planck Institute for Human Development}, \orgaddress{\city{Berlin}, \country{Germany}}}

\address[4]{\orgdiv{Berlin Center for Advanced Neuroimaging}, \orgname{Charité -- Universitätsmedizin Berlin, Corporate Member of Freie Universität Berlin, Humboldt-Universität zu Berlin, and Berlin Institute of Health}, \orgaddress{\city{Berlin}, \country{Germany}}}

\corres{Felix Frederik Zimmermann, Physikalisch-Technische Bundesanstalt (PTB), Abbestrasse 2-12, 10587 Berlin, Germany \email{felix.zimmermann@ptb.de}}

\fundingInfo{Metrology for Artificial Intelligence for Medicine (M4AIM) Project that is funded by the German Federal Ministry for Economic Affairs and Climate Action (BMWi) in the framework of the QI-Digital initiative and in part by the Deutsche For\-schungs\-gemein\-schaft (DFG, German Research Foundation) under Grant Nos. 289347353 (GRK2260, BIOQIC) and 372486779 (SFB1340), and in part by the project 22HLT02 A4IM which has received funding from the European Partnership on Metrology, co-financed from the European Union’s Horizon Europe Research and Innovation Programme and by the Participating States.}

\abstract[Abstract]{\rev{\textbf{Purpose} To provide a reproducible framework for MR image reconstruction and quantitative parameter estimation, with particular relevance to low-field MRI.}
\textbf{Methods} MRpro is an open-source image reconstruction package built upon PyTorch, enabling modern deep-learning reconstructions. It uses open data formats for input \rev{and output (ISMRMRD, DICOM, NIfTI)}, allowing easy integration into existing pipelines and support for data from different devices.
The framework comprises three main areas. First, it provides unified data structures for the consistent manipulation of MR datasets and their associated metadata (e.g., k-space trajectories). Second, it offers a library of composable operators, proximable functionals, and optimization algorithms, including a unified Fourier operator for all common trajectories and operators specifically developed for low-field applications, such as a B$_0$-correction operator. These components are used to create ready-to-use implementations of key reconstruction algorithms. Third, for deep learning, MRpro includes essential building blocks such as data-consistency layers, differentiable optimization layers, state-of-the-art backbone networks, and access to public datasets to facilitate reproducibility.
MRpro is developed as a collaborative project supported by automated quality control.

\rev{\textbf{Results} We demonstrate MRpro across automatic reconstruction, iterative SENSE, deep-learning-based reconstruction, and quantitative parameter estimation. For the selected Cartesian and non-Cartesian tasks, iterative SENSE reconstructions from MRpro differed from those obtained with BART, SigPy, and MRIReco by less than 3\% relative RMSE. Further applications use public and simulated datasets and measured low-field data acquired at 0.3\,T, 0.6\,T, and 47\,mT.}

\rev{\textbf{Conclusion} MRpro provides a general framework for MR image reconstruction and quantitative parameter estimation, with particular relevance to low-field applications.} With reproducibility and maintainability at its core, it facilitates collaborative development and provides a foundation for future MR imaging research.
}

\maketitle

\newskip\oldfootins
\oldfootins=\skip\footins
\global\skip\footins=-.5cm
\renewcommand\thefootnote{}
\renewcommand\thefootnote{}
\footnotetext{* These authors contributed equally.\vspace{1cm}}
\footnotetext{
\textbf{Abbreviations:} MRI, Magnetic Resonance Imaging; LF, Low Field; SNR, Signal-to-Noise Ratio; DL, Deep Learning; TV, Total Variation; DWI, Diffusion-Weighted Imaging; ADC, Apparent diffusion coefficient; SE, Spin Echo; MESE, Multi-echo Spin Echo; SSIM, Structural Similarity Index; MSE, Mean Squared Error; FFT, Fast Fourier Transform; NUFFT, Non-uniform Fast Fourier Transform; CG, Conjugate Gradient; PDHG, Primal-Dual Hybrid Gradient; qMRI, quantitative MRI; EPG, Extended Phase Graph
}
\renewcommand\thefootnote{\fnsymbol{footnote}}
\setcounter{footnote}{1}

\section{Introduction}\label{sec:intro}
\afterpage{\global\skip\footins=\oldfootins}
Magnetic Resonance Imaging (MRI) is a major diagnostic tool in clinical practice and is typically performed at field strengths of $\ge$ \qty{1.5}{\tesla}. Recently, interest in \rev{low-field (LF) systems has increased. Lower field strengths } enable cheaper and more portable hardware by reducing weight, cooling demands, and power consumption. LF MRI therefore has the potential to improve affordability and access in resource-limited and remote settings and to enable point-of-care MRI \cite{Arnold2023,hennig2023evolution,physics}. In addition, lower specific absorption rates (SAR) and reduced susceptibility artifacts can improve safety and image quality in the presence of metallic implants and surgical equipment.

However, image reconstruction faces substantial hardware-related challenges. Lower field strengths inherently reduce the signal-to-noise ratio (SNR)\rev{\cite{edelsteinIntrinsicSignaltonoiseRatio1986}}. Cost-effective magnet designs often lead to stronger B$_0$ inhomogeneities, and the relatively flat sensitivity profiles of receiver coils limit the effectiveness of conventional multi-coil parallel imaging.

Advanced reconstruction methods such as compressed sensing \cite{lustig2007} and deep learning (DL) approaches \cite{Chen2022x,heckel2024deep,aggarwal2018modl,sriram2020end} were originally developed largely for accelerated conventional MRI. These methods are now being adapted to LF imaging to mitigate its physical limitations \cite{Ayde2025,Laguna2022,koonjoo2021boosting}.

A major challenge for advanced and learned reconstruction methods is reproducibility and comparability. Even when code is publicly available, methods often depend on dedicated MR sequences that are not readily accessible, custom raw k-space formats, or highly specific acquisition settings. Long-term maintenance and bug fixes are also often lacking. This makes it difficult to verify results, compare methods, and build on prior work.

Two important milestones for standardizing MR imaging were the introduction of vendor-independent, open-source formats for raw MR data (ISMRMRD) \cite{Inati2017} and MR sequences (Pulseq) \cite{Layton2017}. Pulseq provides a flexible open-source framework for sequence programming, and its format is also supported by closed-source alternatives such as gammaSTAR \cite{konstantin_gammaSTAR_2025}. Many open-source packages for scanner control, including MARCOS \cite{NEGNEVITSKY2023107424} and NEXUS \cite{schote2025}, use Pulseq to define sequences.

A third milestone is the development of open-source reconstruction software. Prominent examples are Gadgetron \cite{gadetron} and the Berkeley Advanced Reconstruction Toolbox (BART) \cite{bart}. Gadgetron is designed primarily for rapid processing of incoming data streams and online reconstruction, whereas BART mainly targets offline reconstruction and method development.
A key limitation is that both frameworks are implemented in C or \CXX. This improves computational performance but complicates integration with ML frameworks such as PyTorch \cite{pytorch} or JAX \cite{jax}. BART partly addresses this by providing custom automatic differentiation functionality \cite{bart}. However, this increases maintenance burden, as new ML building blocks must be reimplemented within the reconstruction framework. As a result, although Gadgetron and BART provide MATLAB and Python interfaces, extending their core functionality still requires substantial C or \CXX\ expertise. This is a barrier for researchers working in the Python-based machine learning ecosystem.

As LF MRI develops rapidly, with commercial products emerging and many groups pursuing custom systems \cite{winter2024open,hennig2023evolution}, the need for a reliable, modern, open-source framework for advanced reconstruction is increasing. Such a framework must address the particularities of LF research systems, including B$_0$ inhomogeneities that break the simple Fourier relationship between acquired data and image space, low SNR, and custom sequences and trajectories. It must also support rapid prototyping of new reconstruction algorithms and DL methods.

These methods often require retrospective undersampling of k-space data. For standard Cartesian acquisitions, this is straightforward, but for non-Cartesian sequences it can be challenging to keep k-space data, trajectory information, and acquisition metadata, such as timing information required for quantitative imaging, consistent. Similarly, combining data from multiple scans, for example to exploit redundancy across contrasts in advanced quantitative reconstructions, while avoiding repeated storage of trajectories and metadata, requires dedicated data handling.

Here we present \texttt{MRpro}, a PyTorch-based, open-source framework for MR image reconstruction and quantitative parameter estimation. The key features of \texttt{MRpro} include:

\begin{itemize}
\itemsep-0.1em
\item implementation in PyTorch for seamless compatibility with ML methods.
\item k-space trajectory calculation from Pulseq \cite{Layton2017} files, with automatic configuration of the corresponding Fourier operators to minimize additional user input.
\item modular design for building complex reconstruction pipelines in Python by combining predefined operators and algorithms.
\item dedicated data structures for unified handling of k-space data, sampling trajectories, and acquisition metadata, enabling retrospective undersampling, data concatenation, and rearrangement operations.
\end{itemize}

\rev{\texttt{MRpro} provides a unified, PyTorch-native software infrastructure for implementing established reconstruction and qMRI methods as reproducible, differentiable workflows.}
\texttt{MRpro} supports input and output in standardized ISMRMRD \cite{Inati2017} and DICOM formats to ensure compatibility with existing processing pipelines. Automatic quality control and continuous integration tests on phantom and in vivo data support robust community-driven development.

\section{Methods}
\begin{figure*}[t]
  \centering\includegraphics[width=0.99\linewidth]{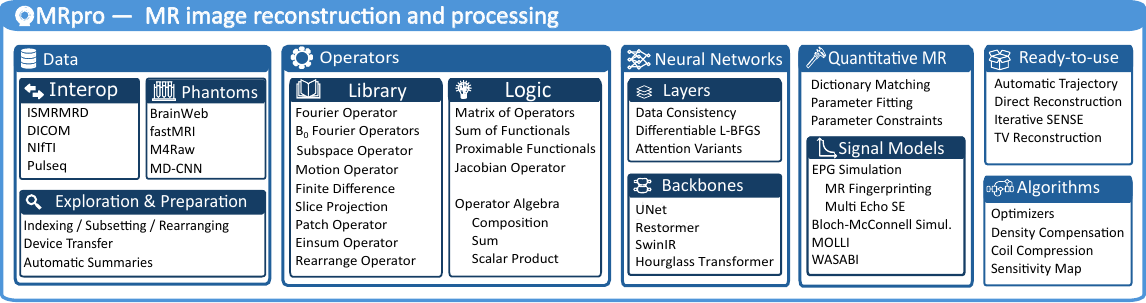}
    \caption{High-level overview of \texttt{MRpro}, covering data handling, a library of composable operators, DL layers and common backbone networks, algorithms, and signal models for qMRI. For more details, see the main text.}\label{fig:overview}
\end{figure*}

The following provides an overview of the software design and highlights key aspects of \texttt{MRpro} (\autoref{fig:overview}). For more detailed information, we refer to the documentation at \url{https://docs.mrpro.rocks}. \rev{The applications below combine \texttt{MRpro}'s reusable components in reconstruction and quantitative imaging workflows.}

The package \texttt{MRpro} is structured around three main thematic blocks: data handling, mathematical tools, and neural networks.

\begin{figure}[t]
\centering
\includegraphics[width=\linewidth]{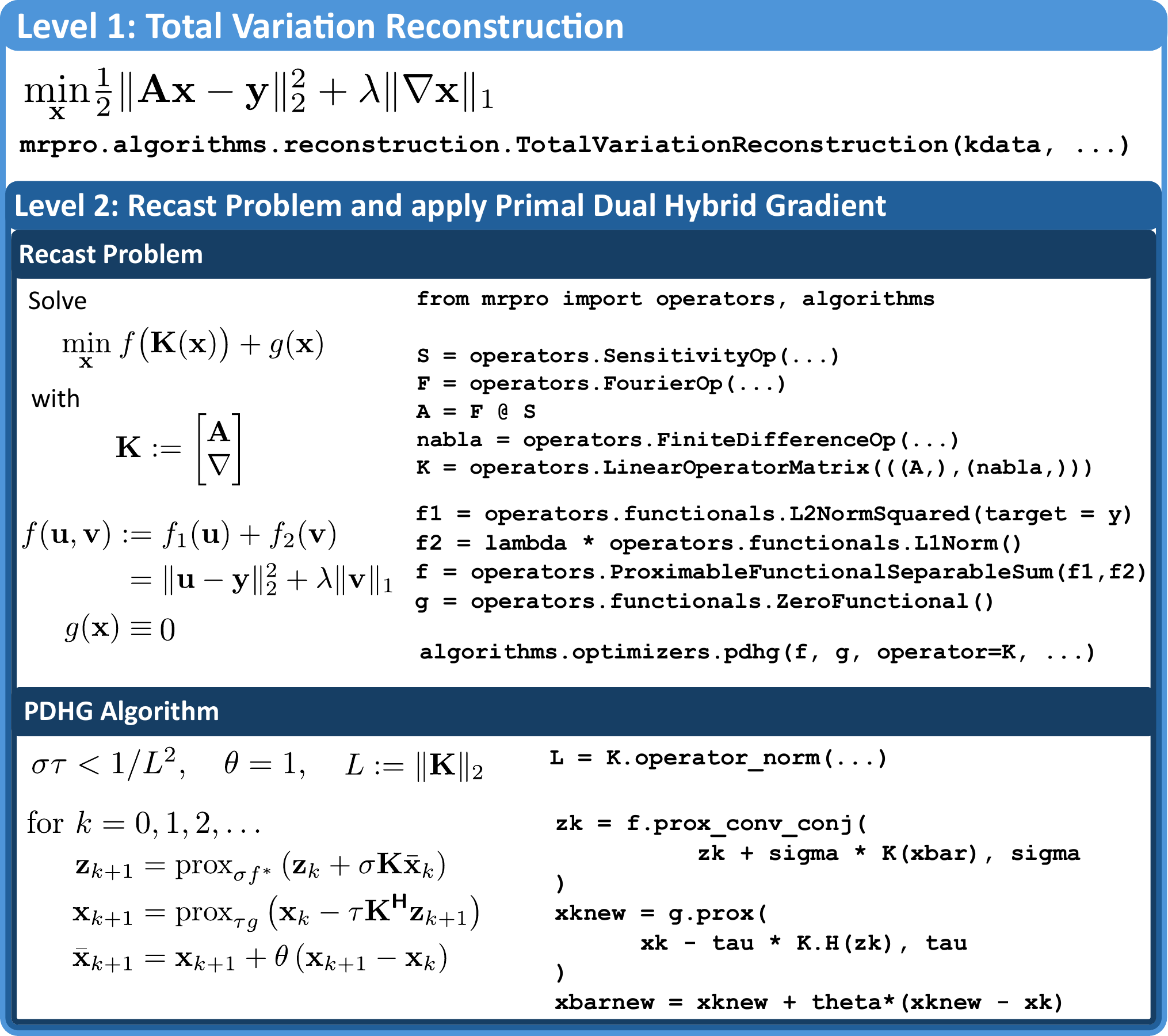}
\caption{Software structure showing the levels of abstraction for components of a TV-minimization-based image reconstruction. At the highest level (Level 1), the user applies TV reconstruction to a \texttt{KData} object. At Level 2, the problem is reformulated for PDHG, and the required operators and functionals are constructed. Finally, PDHG computes the norm of the stacked operator $\op{K}$ and evaluates the proximal operators of the functionals and their convex conjugates.}\label{fig:math_tv}
\end{figure}

\subsection{Data handling}

We focus on open data formats. \rev{Raw k-space data can be imported and exported in ISMRMRD format. Reconstructed image data or quantitative parameter maps can be imported using DICOM. Export of images or parameter maps is possible as DICOM (magnitude only) or NIfTI (magnitude or complex-valued).} Trajectory information can also be imported from ISMRMRD files or directly from Pulseq \cite{Layton2017} sequences.
\texttt{MRpro} utilizes specialized container classes to store k-space data, quantitative parameter maps, and images. These containers hold both the data and its associated metadata, including acquisition information, orientation, position, and k-space trajectories. This design ensures coherent data manipulation; for example, selecting a subset of a \texttt{KData} object along the phase-encoding dimension also subsets the corresponding trajectory and acquisition metadata.

All data containers can be moved to GPU memory for accelerated processing. Additionally, the framework provides human-readable descriptions and summary statistics to facilitate rapid data exploration and preprocessing.

\subsection{Mathematical Framework}
A core component of \texttt{MRpro} is a framework of composable mathematical operators and functionals.
More precisely, \texttt{MRpro} treats operators as mappings between discretized function spaces, represented as tensors of images or measurements. We provide the logic required for arithmetic operations between operators, for example the sum of linear operators $(\op{A}+\op{B})(\vec{x}):=\op{A}\vec{x} + \op{B}\vec{x}$, their composition $(\op{A}\circ\op{B})(\vec{x}):=\op{A}\op{B}\vec{x}$, stacked operators $[\op{A}, \op{B}]^\trans \vec{x}:=[\op{A}\vec{x}, \op{B} \vec{x}]^\trans$ etc.
We further provide an extensive collection of MR-relevant operators. These include a general Fourier acquisition operator (for Cartesian and non-Cartesian trajectories) with different approaches to off-resonance handling, a motion operator (for motion-corrected image reconstruction), a volume-to-slice projection operator (for slice to volume registration), and a patch-extraction operator (for dictionary learning-based approaches \cite{dictionarylearning}). These are implemented as instances of a \texttt{LinearOperator} class and have associated adjoint/Hermitian operators.
Functionals are scalar-valued operators typically used as loss or objective functions. For proper lower-semicontinuous convex functionals, including the squared $\ell_2$-norm, the $\ell_1$-norm, the $\ell_{2,1}$ norm, and the nuclear norm, \texttt{MRpro} implements their proximal operators and those of their convex conjugates. All of these components can be chained, stacked, and differentiated.

These modules are leveraged to implement several important algorithms. These include the Conjugate Gradient (CG) method for solving linear systems $\op{M}\vec{x} = \vec{b}$ with a symmetric, positive-definite $\op{M}$, as required for iterative SENSE \cite{pruessmann1999, pruessmann2001advances} or within alternating minimization schemes \cite{admm}. The package also implements accelerated proximal gradient methods (FISTA/ISTA) for applications such as wavelet-based compressed sensing \cite{daubechies2004iterative, lustig2007} and the Primal-Dual Hybrid Gradient (PDHG) algorithm for TV-regularized reconstruction \cite{chambolle2011first, knoll2010fast}.

Additionally, \texttt{MRpro} includes implementations of other common MR-specific algorithms, such as coil sensitivity estimation \cite{inati_fast_2014,walsh_adaptive_2000}, density compensation, and coil compression \cite{huang2008software}.

\subsection{Integration with DL}\label{subsec:integration_with_dl}
Basing \texttt{MRpro} on PyTorch provides native automatic differentiation and GPU acceleration for all operators and algorithms and enables seamless integration with the broader ML ecosystem.
We provide implementations of modern image-to-image networks, such as U-Net \cite{ronneberger2015u}, Attention U-Net \cite{attentionunet}, Restormer \cite{restormer}, Uformer \cite{uformer}, SwinIR \cite{swinir}, and Hourglass transformer \cite{hourglass}. These architectures are adapted to accept additional conditioning inputs \cite{film}, allowing them to serve as backbones in MR reconstruction networks.
In addition to the building blocks required for these networks, such as various attention mechanisms, \texttt{MRpro} includes multiple data-consistency layers for physics-informed learning. These include a gradient-descent layer, an analytic data-consistency layer for single-coil Cartesian data \cite{schlemper_deep_2018}, and a CG-based layer for multi-coil or non-Cartesian imaging \cite{aggarwal2018modl}. For quantitative MR (qMRI), we implement a differentiable L-BFGS solver \cite{shannoConditioningQuasiNewtonMethods1970} based on implicit differentiation \cite{optnet,pinqi} to enforce consistency with the signal model.

Furthermore, we integrate several well-known public datasets for training and evaluation, including M4Raw \cite{m4raw}, fastMRI \cite{zbontar2018fastmri}, BrainWeb \cite{brainweb}, and the MD-CNN dataset \cite{mdcnn}.

\subsection{Automatic reconstruction}
The main building block of any MR image reconstruction is the Fourier transform. The type of Fourier transform depends on the k-space trajectory, e.g., FFT for Cartesian acquisitions and NUFFT for non-Cartesian acquisitions such as radial or spiral. For a proper reconstruction, detailed knowledge about the actual implementation of the k-space trajectory needs to be provided to the reconstruction algorithm.

In \texttt{MRpro}, all information describing advanced Cartesian acquisition schemes (undersampling, partial Fourier, reversed readouts, etc.) is imported from ISMRMRD. For non-Cartesian acquisition schemes, some
commonly used trajectories such as radial sampling are implemented. Rather than requiring custom sampling schemes to be implemented in \texttt{MRpro}, trajectory information can also be provided as part of the ISMRMRD raw data file. Finally, \texttt{MRpro} also allows for the automatic calculation of the trajectory from a provided Pulseq sequence file.
Based on the trajectory information stored in the \texttt{KData} instance, a Fourier operator can be instantiated. The required Fourier operations are automatically selected, i.e., the dimensions along which FFT or NUFFT operations have to be applied are identified automatically, without user input. Furthermore, a matching density compensation function is calculated. This means that data obtained with arbitrary trajectories can be reconstructed from an MR raw data file and the corresponding Pulseq sequence file without any additional inputs.
This is especially useful for researchers who focus on reconstruction algorithm development and are less experienced with MR data acquisition.
\rev{Multi-coil image data can be combined using root-sum-of-squares, which returns a magnitude image, or sensitivity-map weighting, which preserves complex image information. Sensitivity maps can be provided from calibration measurements or estimated using the Walsh or Inati methods \cite{walsh_adaptive_2000,inati_fast_2014}.}

Furthermore, we provide \rev{complete ready-to-use high-level end-to-end reconstruction pipelines for direct reconstruction with the adjoint acquisition operator, regularized iterative SENSE, and PDHG-based TV-regularized reconstructions}. These include all necessary steps to obtain images from the imported \texttt{KData} alone. In turn, this allows researchers focused on sequence development to perform reproducible reconstructions without requiring a detailed understanding of reconstruction algorithms.

\subsection{Quantitative parameter estimation}\label{subsec:quantitative_parameter_estimation}
\texttt{MRpro} provides several differentiable MR signal models. These include simple mono-exponential T$_1$ or T$_2$ models, Modified Look-Locker Inversion recovery (MOLLI) \cite{Messroghli2004}, Watershift and B$_1$ mapping (WASABI) \cite{schuenkeSimultaneousMappingWater2017}, extended phase graph (EPG) simulations \cite{Weigel2015} (for example, for MR fingerprinting \cite{ma2013,Schuenke2025}), \rev{and Bloch--McConnell simulations}. The models calculate signals from multi-dimensional parameter inputs and are defined as (non-linear) \texttt{Operator} objects that can be readily combined with other operators. In order to fit a T$_1$ inversion-recovery model to magnitude images, for example, the signal model operator can be combined with the magnitude operator to form a new model.

In addition to non-linear optimizers \cite{kingmaAdamMethodStochastic2015a,shannoConditioningQuasiNewtonMethods1970} for parameter regression, \texttt{MRpro} includes a dictionary-matching operator (see \autoref{subsec:mrfmethod}).

\subsection{Community-based development, quality control and release}
Built in Python and licensed under the Apache 2.0 license, \texttt{MRpro} is developed openly on GitHub to foster community collaboration and reproducibility.

\begin{figure*}[t]
  \centering
  \includegraphics[width=\textwidth]{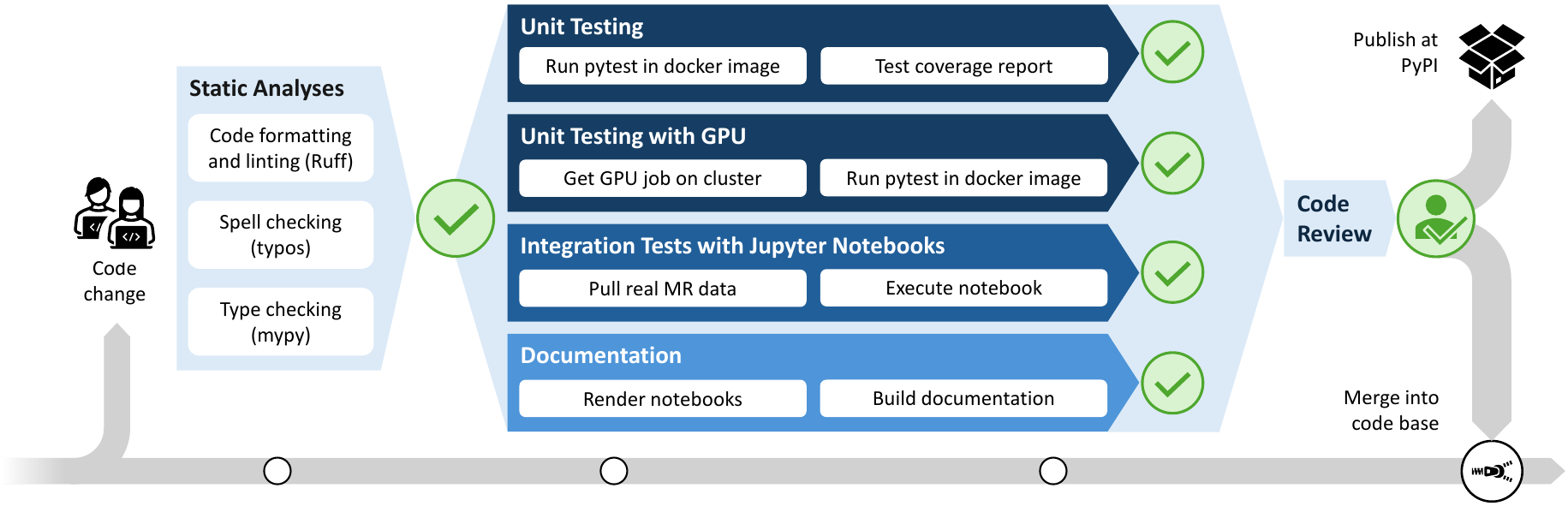}
  \caption{
      Flowchart of the CI workflow for a pull request. Static checks, unit tests, and integration tests are executed. The integration tests are notebooks and also serve as examples in the documentation. After a manual code review, the change is merged.
  }
  \label{fig:ci}
\end{figure*}

High code quality, consistent coding practices, and adherence to established standards are enforced through a continuous integration (CI) pipeline, illustrated in \autoref{fig:ci}. This pipeline comprises automated static analysis (Ruff\footnote{\url{https://github.com/astral-sh/ruff}}, Mypy\footnote{\url{https://github.com/python/mypy}}) and a multilevel testing strategy. The unit tests verify individual components and achieve more than \qty{90}{\percent} code coverage. The tests are run partly on GitHub servers, partly on GPU-accelerated nodes within our in-house compute cluster. Integration tests validate end-to-end functionality on realistic MRI reconstruction tasks and are subsequently converted into Jupyter notebooks, serving as executable examples within the documentation. \rev{The integration tests verify, for example, that reconstruction algorithms converge to the same solution or that estimated quantitative parameters match known reference values.} Before a proposed code change is merged into the codebase, all tests must pass, and changes to the documentation and the code are reviewed.
We provide regular releases on both GitHub and PyPI. For backward compatibility, we follow the \textit{Scientific Python}\footnote{\url{https://scientific-python.org/}} recommendations and support the last three years of Python versions.
We provide full Docker images for reproducibility as well as Docker recipes for use with the Siemens OpenRecon system for online reconstruction\footnote{\url{https://github.com/fzimmermann89/mrpro_server}}.

\subsection{Experiments}

In the following, we give an overview of image reconstruction and parameter estimation examples. If not stated otherwise, we consider variational reconstruction methods of the form
\begin{equation}\label{eq:variational_recon_problem}
    \underset{\vec{x}}{\min} \frac{1}{2}\| \op{A}\vec{x} - \vec{y}\|_2^2 + \mathcal{R}(\vec{x}),
\end{equation}
where $\op{A}$ denotes the linear forward operator (usually the composition of a Fourier operator and a coil-sensitivity operator), $\vec{x}$ denotes the sought complex-valued MR image, $\vec{y}$ denotes the corresponding (often undersampled and noise-corrupted) k-space data, and $\mathcal{R}$ denotes an optional regularization functional for the image.
We also consider variational quantitative parametric estimation problems of the form
\begin{equation}\label{eq:variational_qrecon_problem}
    \underset{\vec{p}}{\min} \frac{1}{2}\| \op{A}\op{q}(\vec{p}) - \vec{y}\|_2^2 + \mathcal{S}(\vec{p}),
\end{equation}
    where $\op{q}$ denotes a non-linear voxel-wise signal model (e.g., inversion recovery, saturation recovery), $\vec{p}$ denotes the quantitative parameter maps of interest (e.g., M$_{0}$, T$_{1}$, T$_{2}$), and $\mathcal{S}$ denotes a regularization functional for the parameter maps.

\rev{Data acquired at PTB were obtained on either a Siemens Cima.X system at \qty{3}{\tesla} or an OSI$^2$ ONE system at \qty{47}{\milli\tesla}. Additional data were drawn from public datasets \cite{brainweb,zbontar2018fastmri,ibt_cmr_data,m4raw,osi2_leiden_data}.}
\rev{The \qty{47}{\milli\tesla} data were acquired on two separate OSI$^2$ ONE scanners, one at PTB and one in Leiden.}

Public datasets, simulations, \rev{and \qty{3}{\tesla} data were used for reproducible benchmarking and general framework demonstrations, whereas measured low-field data were used to evaluate low-field applications.} All in vivo scans were approved by the local ethics committees and were carried out in accordance with relevant guidelines and regulations. The subjects gave written informed consent to participate in this study.

\subsubsection{Automatic reconstruction and data handling}

In the first experiment, we demonstrate three different approaches for defining or extracting the k-space trajectory within the reconstruction pipeline of a 2D radial gradient-echo dataset (402 spokes, golden-angle) acquired on the 3\,T Siemens Cima.X system. These approaches include i) utilizing the built-in analytical 2D radial trajectory calculator (\texttt{KTrajectoryRadial2D}), ii) extracting the trajectory information embedded within the ISMRMRD raw data file (\texttt{KTrajectoryIsmrmrd}), and iii) automatically calculating the trajectory from the corresponding Pulseq sequence file (\texttt{KTrajectoryPulseq}).

In all cases, images were generated using \texttt{MRpro's} high-level \texttt{DirectReconstruction} class, which automatically creates the required Fourier and density compensation operators, as well as coil sensitivity maps based solely on the provided \texttt{KData} object.

In addition, to demonstrate the indexing and subsetting capability of \texttt{MRpro}, a subset (the last 40 spokes) of the data was extracted using standard Python indexing syntax on the \texttt{KData} object. This operation inherently subsets the k-space data along with the corresponding k-space trajectory and acquisition metadata.

\rev{
\subsubsection{Comparison to other reconstruction frameworks}

We compared \texttt{MRpro} to three other commonly used image reconstruction frameworks:
\begin{itemize}
\item Berkeley Advanced Reconstruction Toolbox (BART) (version 1.0.00, written in C) \cite{bart}
\item MRIReco (version 0.9.2, written in Julia) \cite{mrireco}
\item SigPy (version 0.1.27, written in Python) \cite{sigpy}
\end{itemize}

We selected iterative SENSE \cite{pruessmann2001advances} because it was implemented in all four frameworks. Two different datasets were used for evaluation. The first one was a 3D Cartesian acquisition of the brain obtained from an OSI$^2$ ONE open-source\footnote{\url{https://www.opensourceimaging.org/project/osii-one/}} ultra-low-field scanner \cite{oreilly,winter2024open} (T$_1$-weighted turbo spin-echo sequence, echo train length 8, echo time \qty{20}{\milli\second}, repetition time \qty{1200}{\milli\second}, field of view \qtyproduct{225x205x190}{\milli\meter\cubed}, resolution \qtyproduct{1.5x1.5x5}{\milli\meter\cubed}, and a partial Fourier factor of 0.65; a T$_1$ inversion-preparation pulse with an inversion time of \qty{90}{\milli\second} was applied to increase T$_1$ contrast). The second was a 2D golden-radial cardiac acquisition obtained with a Siemens Cima.X \qty{3}{\tesla} scanner (gradient-echo sequence, echo time \qty{1.7}{\milli\second}, repetition time \qty{3.4}{\milli\second}, field of view \qtyproduct{320x320}{\milli\meter\squared}, in-plane resolution \qtyproduct{2x2}{\milli\meter\squared}, and slice thickness \qty{8}{\milli\meter}). For the Cartesian dataset, all frameworks used one iterative SENSE iteration. For the non-Cartesian dataset, 20 iterations were used for \texttt{MRpro}, BART, and SigPy, whereas five iterations
  were used for MRIReco because its iterative SENSE implementation applies trajectory-based density-compensation weighting, resulting in different convergence behavior.

    The same coil sensitivity maps were used for all reconstruction frameworks. We calculated the relative RMSE between each \texttt{MRpro} reconstruction and the corresponding reconstructions from the other frameworks as a measure of cross-framework reconstruction accuracy for the selected tasks. Code for data preparation, reconstruction, and plotting is available online.\footnote{\href{https://github.com/ckolbPTB/OpenSourceMrRecon/tree/mrpro}{\nolinkurl{github.com/ckolbPTB/OpenSourceMrRecon}}}
}

\subsubsection{Learning spatially adaptive regularization parameter maps for TV-based reconstruction}

In TV-based reconstruction, using a single regularization parameter that globally dictates the strength of the TV regularization, i.e., $\mathcal{R}(\vec{x}) = \lambda \| \nabla \vec{x}\|_1$ with $\lambda>0$ in \autoref{eq:variational_recon_problem}, is suboptimal. Instead, locally adapting the regularization strength by employing parameter maps can substantially improve the resulting reconstructions \cite{pragliola2023and}.
\rev{We implemented the learned spatially adaptive TV method described in \cite{kofler2023learning} using \texttt{MRpro}'s differentiable operators and PDHG solver.}
The approach in \cite{kofler2023learning} consists of two main blocks. First, a convolutional neural network (CNN), $\mathcal{N}_\Theta$, is applied to an initial estimate of the image $\vec{x}_0$ to estimate suitable regularization parameter maps, i.e., $\Lambda_{\Theta}:=\mathcal{N}_\Theta(\vec{x}_0)$. Second, an unrolled PDHG scheme of finite length solves the problem
\begin{equation}\label{eq:tv_problem_with_lambda_map}
    \underset{\vec{x}}{\min}\, \frac{1}{2}\|\op{A}\vec{x} - \vec{y}\|_2^2 + \|\Lambda_{\Theta}\nabla \vec{x}\|_1 \vspace{0.1cm}
\end{equation}
assuming the regularization parameter maps are fixed. The entire reconstruction network can be trained end-to-end; thus, $\mathcal{N}_{\Theta}$ learns to estimate regularization parameter maps such that the resulting reconstructions are as close as possible to the target images.
\rev{We provide $\mathcal{N}_\Theta(\vec{x}_0)$ as the \texttt{weight} parameter for the functional $f_2$ in \autoref{fig:math_tv} to define a weighted \(\ell_1\)-norm.}

We trained the network on simulated data based on the fastMRI dataset \cite{zbontar2018fastmri}.
\rev{We then applied the trained network to two publicly available T$_2$-weighted low-field datasets: the IBT-CMR dataset \cite{ibt_cmr_data}, acquired with a Philips Ingenia Ambition X scanner ramped down
  to \qty{0.6}{\tesla}, and the OSI$^2$ dataset \cite{osi2_leiden_data}, acquired at \qty{47}{\milli\tesla} with an OSI$^2$ ONE scanner operated in an intensive care unit in Leiden.}

\subsubsection{Diffusion-weighted image reconstruction} \label{subsec:diffusion}
Diffusion-weighted imaging (DWI) is sensitive to the microscopic mobility of water in tissue and can probe diffusion restriction associated with many diseases such as acute ischemic stroke or tumor formation. However, DWI usually requires prolonged echo times to accommodate diffusion-sensitive gradients, which reduce SNR. This becomes especially challenging for low-SNR applications, such as low-field MRI. Therefore, we demonstrate the application of \texttt{MRpro} image reconstruction and apparent diffusion coefficient (ADC) parameter estimation in noise-prone DWI acquisitions on the \rev{\qty{47}{\milli\tesla} OSI$^2$ ONE scanner at PTB} using a commercial phantom
(CaliberMRI: Low Field Single Plate Phantom Model 139, serial number: 139-0003).

Four of the spherical inclusions in the central section of the phantom, each \qty{17}{\milli\meter} in diameter, had prescribed ADC values of \qtylist{1.591;1.211;0.876;0.527}{\micro\meter\squared\per\milli\second} at \qty{21}{\degreeCelsius} and were used for ADC mapping. DWI was performed using a series of turbo spin-echo sequences with monopolar diffusion-encoding gradients of varying strength to achieve b-values of \qtylist{0;41;111;208}{\second\per\milli\meter\squared}. Additional imaging parameters were an isotropic resolution of \qty{2.8}{\milli\meter}, a field of view of \qty{200}{\milli\meter}, a repetition time of \qty{1000}{\milli\second}, an echo time of \qty{10}{\milli\second}, and an echo train length of 10. Diffusion encoding was performed simultaneously along the scanner's y- and z-axes. Images were reconstructed in \texttt{MRpro} using adjoint reconstruction and, for comparison, iterative TV-regularized reconstruction (see also \autoref{fig:math_tv}). Reconstructed images were used to calculate voxel-wise ADC values based on a simple linear regression of the logarithmic signal amplitude versus b-values. ADC values from the central slice were averaged for each inclusion and compared to the provided reference values. The analysis regions contained 25 voxels on average.

The sequence (Pulseq) and the raw acquired data are available on Zenodo \cite{diffusion_zenodo}.

\subsubsection{Complex Signal Models} \label{subsec:mrfmethod}
MR relaxometry such as T$_{2}$-mapping can provide important diagnostic information. In clinical high-field MRI practice, sequences are commonly used that permit simple signal models (e.g., mono-exponential models). \rev{More advanced models such as Extended Phase Graph (EPG) \cite{Weigel2015} are only needed for very specific techniques.} Due to cost-efficient hardware design in low-field MRI, assumptions about signal behavior such as mono-exponential decay are often invalid. Therefore, even for straightforward relaxometry sequences, complex signal models are needed.

Here we demonstrate this using an open-source multi-echo spin-echo (MESE) sequence acquired with a \qty{47}{\milli\tesla} OSI$^2$ ONE LF scanner. Imaging parameters were as follows: field of view, \qtyproduct{140x140}{\milli\meter\squared}; repetition time, \qty{1500}{\milli\second}; MESE echo time, \qty{16}{\milli\second}; and echo train length, 10. For comparison, a spin-echo (SE) sequence with echo times of \qtylist{20;40;80;160;320}{\milli\second} was acquired using otherwise identical sequence parameters. We used a custom phantom holder \cite{gitlab_phantom} consisting of seven gel-filled tubes (Leeds Test Objects Ltd: T$_{1}$/T$_{2}$ EUROSPIN Tubes).
The sequence and acquired data are available on Zenodo \cite{t2_zenodo}.

\rev{For the MESE sequence, each echo image was obtained using adjoint reconstruction. Parameter maps were then estimated by matching each pixel's measured signal evolution to entries in a signal dictionary.} Specifically, for a given pixel's measured signal evolution $\vec{x}$, the index $k^\ast$ of the best-matching dictionary entry $\vec{d}_k$ is found by maximizing the normalized dot product \cite{ma2013},
\begin{equation}
    k^\ast = \underset{k \in \{1, \dots, K\}}{\arg\max}\, \left| \left\langle \frac{\vec{d}_k}{\|\vec{d}_k\|_2}, \frac{\vec{x}}{\|\vec{x}\|_2} \right\rangle \right| .
\end{equation}
The quantitative parameters $\vec{p}_{k^\ast} = (T_2, B_1)_{k^\ast}$ associated with this index were then assigned to the pixel.

In this experiment, we compared $T_2$ maps obtained with a simple mono-exponential signal model and a complex EPG model that also included B$_1$ variation. T$_2$ values between \qty{10}{\milli\second} and \qty{300}{\milli\second} and relative B$_1$ values between 0.6 and 1.0 were considered in the dictionary.

\rev{For the SE sequence, a T$_2$ map was obtained using the same dictionary-matching approach with a mono-exponential signal model and T$_2$ values from \qty{6}{\milli\second} to \qty{500}{\milli\second}.}

\subsubsection{Model-based DL}

In \textit{MoDL} \cite{aggarwal2018modl}, the regularization in \autoref{eq:variational_recon_problem} is learned, i.e.\ $\mathcal{R}=\mathcal{R}_\Theta$
where $\Theta$ denotes the parameters of the CNN $\mathcal{N}_{\Theta}$. The problem is solved by alternating between the CNN for denoising/artifact removal and a data-consistency update.
Specifically, starting with an initial reconstruction $\vec{x}_0$, the updates for $k=1,\ldots,T$ are given by
\begin{align}
\vec{z}_k &:= \mathcal{N}_{\Theta}(\vec{x}_k)
\label{eq:modl_prox_step}
\\
\vec{x}_{k+1} &:= \underset{\vec{x}}{\arg \min} \ \frac{1}{2} \| \op{A} \vec{x} - \vec{y} \|_2^2 + \frac{\lambda}{2} \| \vec{x} - \vec{z}_k \|_2^2, \label{eq:modl_dc_step}
\end{align}
where the learned parameter $\lambda>0$ balances the data consistency and the CNN.

While MoDL was initially proposed for multi-coil high-field data, we consider a typical LF scenario \cite{schote}: a single-coil acquisition with B$_0$ inhomogeneities. Thus, $\op{A}$ is a time-segmented Fourier operator \cite{b0fourier}.
\texttt{MRpro} provides the necessary operators, neural-network integration, and an efficient differentiable solution mapping for solving \autoref{eq:modl_dc_step}, making it easy to implement MoDL with only a few lines of code. We implemented MoDL using our implementation of a U-Net with 3 resolution levels, 3 million parameters, and attention at the lowest resolution level. We shared weights across the five iterations, as in the original publication. We used a CG operator with an explicit implementation of the derivative, thereby avoiding storing intermediate results for back-propagation \cite{aggarwal2018modl, pinqi}. \rev{We applied MoDL to downscaled noisy 2D Cartesian data with simulated random but known B$_0$ maps and Gaussian variable-density phase-encoding undersampling ($R=2$), retaining 20 fully sampled central phase-encoding lines.} \rev{The model was trained without augmentations using AdamW with a learning rate of $10^{-4}$ and a batch size of 4; training stopped after one epoch when the validation loss plateaued.} The loss equally weighted mean squared error and structural similarity index \cite{wang2004image} (SSIM).

\subsubsection{End-to-end training of physics-informed quantitative imaging}

A recent DL method, \textit{PINQI} \cite{pinqi}, uses half-quadratic splitting to solve \autoref{eq:variational_qrecon_problem} by alternating between two subproblems. The first is a linear image reconstruction task
\begin{equation}\label{eq:subproblem_x}
\underset{\vec{x}}{\min} \frac{1}{2} \| \op{A} \vec{x} - \vec{y} \|_2^2 + \frac{\lambda_{\vec{x}}}{2} \left\| \vec{x} - \vec{x}_{\text{reg}} \right\|_2^2 + \frac{\lambda_{\op{q}}}{2} \left\| \op{q}(\vec{p}) - \vec{x} \right\|_2^2
\end{equation}
where $\vec{x}$ denotes the intermediate qualitative images, $\lambda_{\vec{x}}$ and $\lambda_{\op{q}}$ denote the regularization strengths, and $\vec{x}_{\text{reg}}$ denotes an image prior. The second, non-linear, subproblem estimates the quantitative parameters by solving
\begin{equation}\label{eq:subproblem_p}
\underset{\vec{p}}{\min} \frac{\lambda_{\op{q}}}{2}\left \| \op{q}(\vec{p}) - \vec{x} \right\|_2^2 + \frac{\lambda_{\vec{p}}}{2} \left\| \vec{p} - \vec{p}_{\text{reg}} \right\|_2^2.
\end{equation}
Here, $\vec{p}_{\text{reg}}$ denotes a prior for the parameter maps, and $\lambda_{\vec{p}}$ denotes its regularization strength.
In PINQI, a solution to \autoref{eq:variational_qrecon_problem} is found by alternating between the two subproblems. In each iteration $k=1,\ldots,T$, the image and parameter priors are updated by U-Nets as $\vec{x}_{\text{reg},k}=\mathcal{N}_\Theta(\vec{x}_{k-1})$ and $\vec{p}_{\text{reg},k}=\mathcal{P}_\Phi(\vec{x}_k)$, respectively. The network parameters $\Theta$ and $\Phi$ and the regularization strengths are trained end-to-end. This requires differentiable solvers for both subproblems so that gradients can flow to all trainable parameters.
\texttt{MRpro} provides flexible operators for differentiable optimization as well as network backbones to implement PINQI.

\rev{We evaluated this implementation on a simulated saturation-recovery T$_1$-mapping task based on the BrainWeb dataset \cite{brainweb}. For each sample and tissue class, T$_1$ and M$_0$ values were drawn from distributions matching the expected values at ultra-low field strengths \cite{Jordanova2023}.} \rev{The signal model maps the parameter maps $\vec{p} = [\vec{M_0}, \vec{T_1}]^\trans$ to a stack of $S=4$ signals corresponding to saturation times $t_i$ (\qty{0.2}{\second}, \qty{0.5}{\second}, \qty{1}{\second}, and \qty{4}{\second}). Its $i$th component is given by $q_i(\vec{M_0}, \vec{T_1}) = \vec{M_0}(1 - \mathrm{e}^{-t_i/\vec{T_1}})$.}
  \rev{We simulated Cartesian undersampling with $R=6$ by retaining 32 of 192 phase-encoding ($k_y$) lines per contrast while fully sampling the readout direction. The masks had a Gaussian variable-density
  distribution with 12 fully sampled central lines and were generated independently for each contrast and regenerated during training.}
For comparison, we used a classical two-step approach: TV-regularized image reconstruction followed by pixel-wise non-linear least-squares fitting of the signal model.
\rev{PINQI was trained for 40 epochs using AdamW with a batch size of 16 and a OneCycle learning-rate schedule with 10\% warm-up. Peak learning rates were $3 \times 10^{-4}$ for the network parameters and $3 \times 10^{-3}$ for the learned regularization strengths.}

\subsubsection{\rev{Noise2Noise-trained reconstruction of low-field data}}
\label{subsec:m4raw_method}
\rev{

 The M4Raw dataset contains repeated acquisitions of the same subjects, providing independent noisy measurements without a noise-free reference. We used these repeats for repeat-supervised Noise2Noise
  training \cite{lehtinen2018noise2noise} on measured multi-coil data.

  We retained the supplied data partitions and pooled the T$_1$- and T$_2$-weighted data from the public four-channel M4Raw dataset acquired at \qty{0.3}{\tesla} \cite{m4raw}. The training set contained 128 subjects (256 subject--contrast groups; 4,608 slices), the validation set 30 subjects, and the test set 25 subjects (50 subject--contrast groups; 900 slices).
  For each training sample, we randomly selected one of three repetitions as input and a different repetition as an independent noisy target. We registered the target to the input by estimating a subpixel translation from the magnitude images and applying it to the target k-space as a phase ramp. The models operated on complex-valued images and used a magnitude mean-squared-error loss.

  We evaluated the acquired $256\times195$ sampling pattern and retrospective four-fold undersampling. The undersampling mask retained 49 phase-encoding lines, including 24 central ACS lines; the remaining lines followed a Gaussian variable-density distribution. We estimated Walsh coil sensitivities and normalization from the ACS data of the input repetition. Adjoint and PDHG-based TV reconstructions served as baselines; a grid search on the training set selected a scalar TV weight of $\lambda=0.1$ for both sampling patterns.

  We compared spatially adaptive TV reconstruction \cite{kofler2023learning}, implemented with a U-Net and differentiable PDHG, with a residual SwinIR denoiser \cite{swinir} and a three-unroll MoDL model \cite{aggarwal2018modl} with independent U-Nets and implicitly differentiated conjugate-gradient data consistency. We trained separate models for each sampling pattern for 24 epochs using AdamW and a OneCycle learning-rate schedule with 10\% warm-up. Batch sizes were 16, 4, and 8 for adaptive TV, SwinIR, and MoDL, respectively, with peak network learning rates of $10^{-4}$, $2 \times 10^{-4}$, and $2 \times 10^{-4}$. The learned TV scale and MoDL regularization weights used peak learning rates of $10^{-3}$ and $10^{-2}$, respectively. Augmentation comprised superior--inferior flips and translations of up to 10 pixels. For testing, we used repetition 1 as input and the registered average of repetitions 2--6 as a higher-SNR reference. We report brain-masked, slice-wise SSIM and PSNR as mean $\pm$ standard deviation across 25 test subjects after within-subject averaging.}

\section{Results}

\subsection{Basic image reconstruction}

\begin{figure}[t]
\centering
\includegraphics[width=0.8\linewidth]{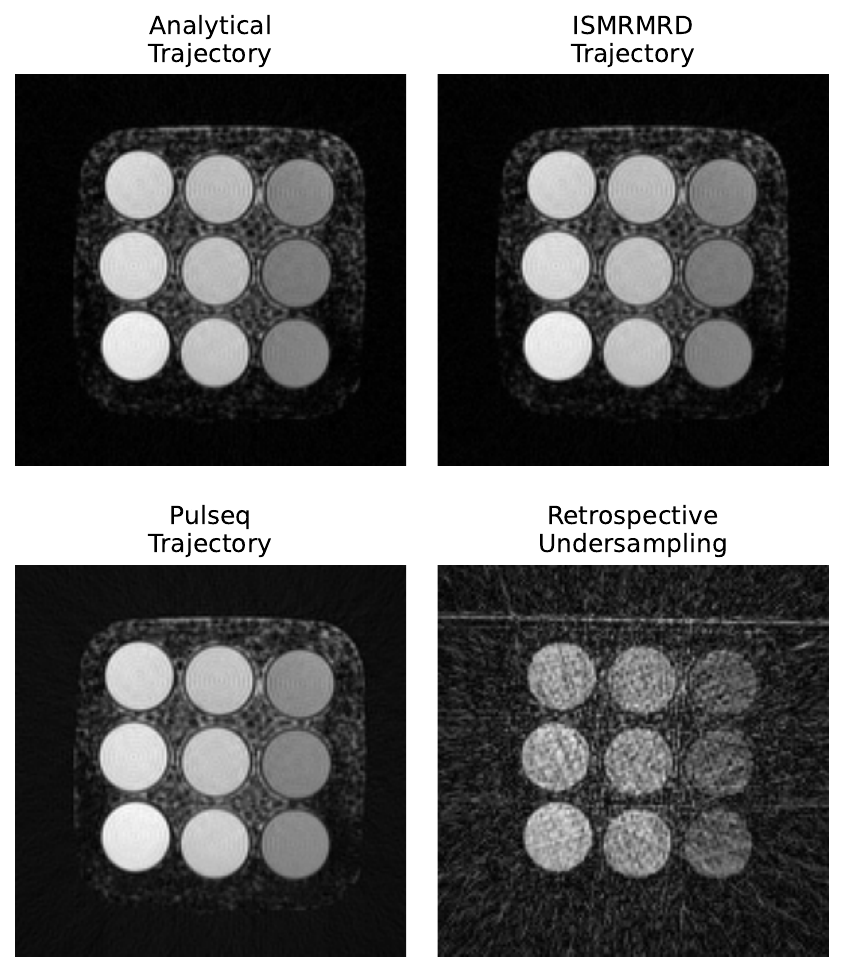}

\caption{Comparison of the automatic reconstruction of radial GRE data (\qty{3}{\tesla} Siemens Cima.X) with trajectory information either (left to right) analytically calculated, read from the ISMRMRD file, or calculated based on the Pulseq sequence. \rev{Visual inspection shows close agreement among the reconstructed images.} Retrospective undersampling (right) can be achieved by subsetting a \texttt{KData} object. This operation automatically subsets the corresponding k-space data, trajectory information, and acquisition headers. }\label{fig:traj}
\end{figure}

\autoref{fig:traj} shows the results of a radial (high-field) image reconstruction using different sources for the trajectory information: an analytical expression, trajectory information read from the ISMRMRD file, and a trajectory calculated based on the Pulseq sequence file.

\rev{Visual inspection showed close agreement among the reconstructed images obtained from the three trajectory sources,} confirming that the different trajectory calculators yield consistent k-space trajectories.

The bottom-right panel of \autoref{fig:traj} shows the results of retrospective undersampling. By subsetting \texttt{KData} to the last 40 spokes, the resulting image exhibits the typical streaking artifacts expected for heavily undersampled radial acquisitions. This confirms that the indexing operation successfully synchronizes the subsetting of the raw data with the corresponding trajectory metadata.

\rev{
\subsection{Comparison to other reconstruction frameworks}
\begin{figure}[t]
\centering
\includegraphics[width=\linewidth]{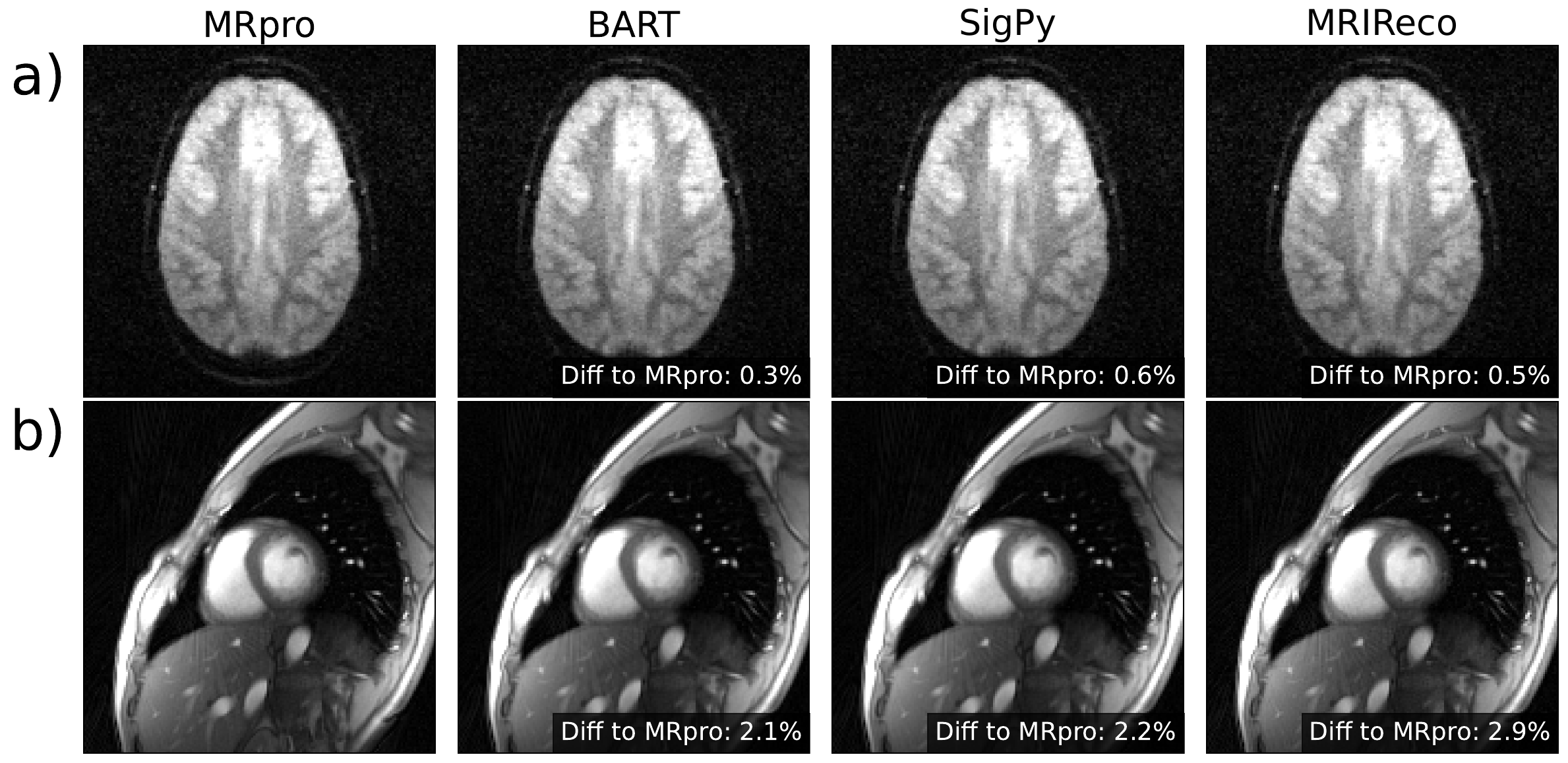}

\caption{Comparison of different image reconstruction frameworks. Image reconstruction is carried out with \texttt{MRpro}, BART, MRIReco and SigPy. a) Reconstruction of a 3D brain scan obtained with a Cartesian sampling pattern. b) Reconstruction of a 2D golden radial acquisition of the heart. Differences between \texttt{MRpro} and the other reconstruction frameworks are reported as relative RMSE values.}\label{fig:compare_frameworks}
\end{figure}

\autoref{fig:compare_frameworks} shows images reconstructed with \texttt{MRpro}, BART, MRIReco and SigPy. The relative RMSE values between \texttt{MRpro} and the other reconstruction frameworks are below \qty{1}{\percent} for the Cartesian reconstruction and below \qty{3}{\percent} for the non-Cartesian data.
}

\subsection{Learning spatially adaptive regularization for TV-based reconstruction}
\label{subsec:tv_maps}
\begin{figure}
\centering
\includegraphics[width=0.98\linewidth]{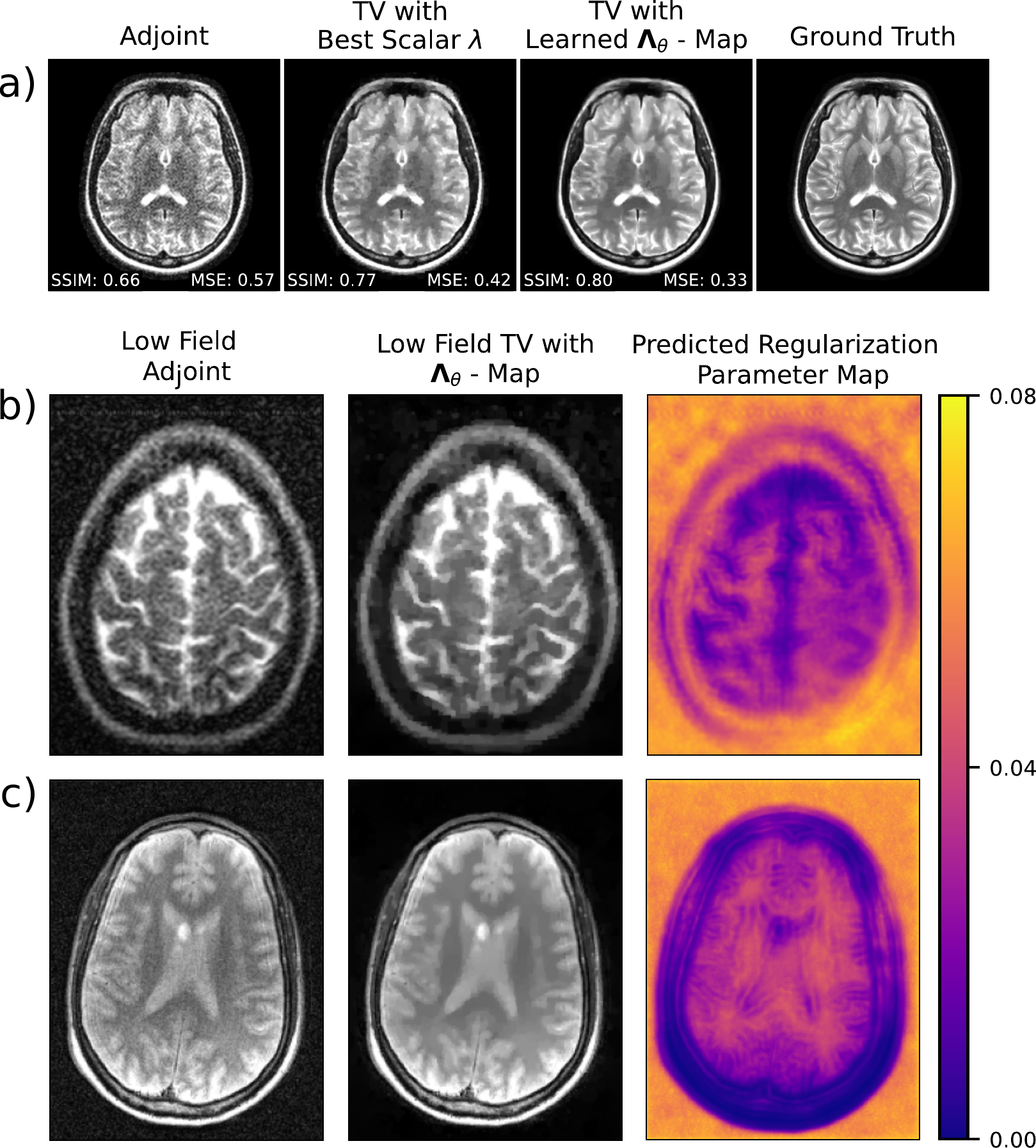}

\caption{Learning spatially adaptive regularization parameter maps for TV-reconstruction. a) Results for a \textit{simulated} LF brain MRI example based on the fastMRI \cite{zbontar2018fastmri} dataset. From left to right: adjoint reconstruction, TV-reconstruction with the best scalar regularization parameter $\lambda$ (obtained by line search using the ground truth image), TV-reconstruction with estimated spatially adaptive regularization parameter maps, and the ground truth image. \rev{b) Application of the network trained on simulated fastMRI data to \textit{in vivo} data using the IBT-CMR data \cite{ibt_cmr_data} (Philips Ingenia Ambition X, ramped down to \qty{0.6}{\tesla}). c) Application to the OSI$^2$ dataset \cite{osi2_leiden_data}, acquired on a \qty{47}{\milli\tesla} OSI$^2$ ONE scanner in Leiden and operated in an intensive care unit.}
}
\label{fig:tv_maps}
\end{figure}

\autoref{fig:tv_maps} shows the results for a learned spatially adaptive TV-based reconstruction. For training and validation, we simulated a noisy low-resolution ($120 \times 120$ matrix) single-coil Cartesian acquisition based on the fastMRI data, using a high-resolution MVUE reconstruction as ground truth.

The adjoint reconstruction is poor (SSIM \num{0.66}, MSE \num{0.57}). While the TV reconstruction with an optimal scalar parameter improves the result (SSIM \num{0.77}, MSE \num{0.42}) and visibly removes noise, the learned spatially adaptive parameter maps yield the best reconstruction (SSIM \num{0.80}, MSE \num{0.33}). Notably, the learned approach is superior even though the scalar baseline was optimized using an oracle line-search with access to the ground truth.

The network trained on simulated data \rev{was also applied to in vivo data acquired on a \qty{0.6}{\tesla} scanner (\autoref{fig:tv_maps}b) and a \qty{47}{\milli\tesla} scanner (\autoref{fig:tv_maps}c). This example showcases the
  implementation of the method \cite{kofler2023learning} using \texttt{MRpro}'s building blocks.}

\subsection{Diffusion-weighted image reconstruction}

\begin{figure}[t]
\centering
\includegraphics[width=\linewidth]{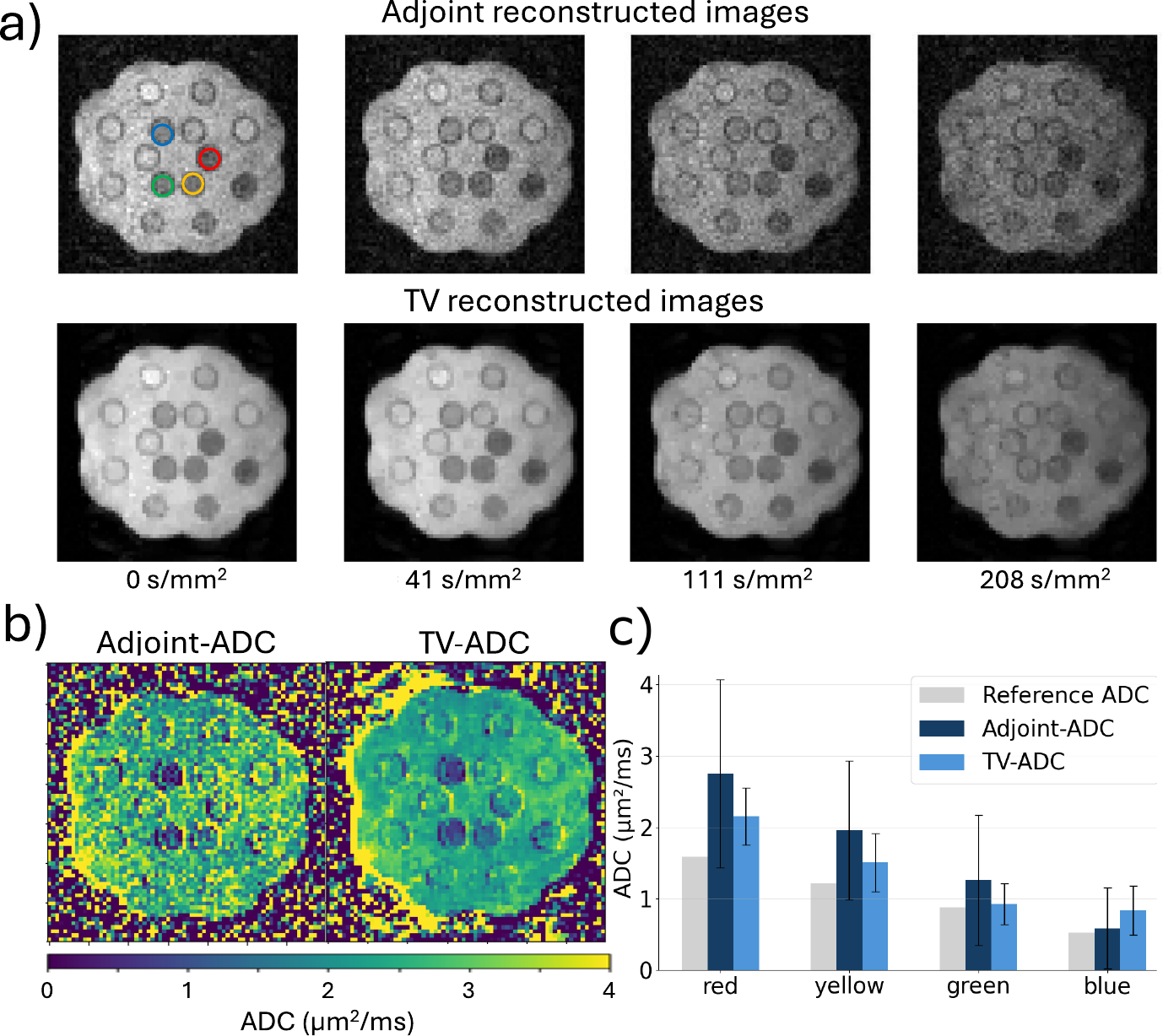}
\caption{Reconstruction of the raw images and ADC maps from a diffusion-weighted TSE sequence. a) Diffusion-weighted images with different b-values reconstructed using adjoint reconstruction (top) and TV-reconstruction (bottom). The colored masks indicate the regions for the ADC comparison to the reference values. b) Maps of ADC values based on adjoint and TV-reconstruction. \rev{c)} Comparison with the vendor-specified ADC values.}\label{fig:dwi}
\end{figure}

\autoref{fig:dwi} displays the results from the DWI and ADC reconstruction. \autoref{fig:dwi}a shows the acquired images at different b-values reconstructed with either the adjoint (top) or the TV-reconstruction (bottom). In addition, colored masks are shown for the numerical comparison of ADC values with reference values. \autoref{fig:dwi}b shows the reconstructed ADC maps, and \autoref{fig:dwi}c compares the estimated ADC values with the vendor-specified reference values. Despite the large voxel volume, the adjoint-reconstructed images exhibited high noise levels, and the SNR decreased further at higher b-values, resulting in noise-dominated ADC maps. TV reconstruction markedly reduced image noise and, consequently, noise in the ADC map. The comparison revealed an overall overestimation of the ADC values, with the estimates from the TV reconstruction more closely matching the vendor-specified values. The maximum absolute differences between the specified and mapped values were \qty{2.4}{\micro\meter\squared\per\milli\second} and \qty{1.2}{\micro\meter\squared\per\milli\second} for the adjoint-based and TV-based reconstructions, respectively. TV reconstruction reduced both the differences from the reference values and the variation within each inclusion, as indicated by the smaller error bars.

\subsection{Complex Signal Models}
\label{subsec:mrf}

\begin{figure}[t]
\centering
{\includegraphics[width=0.95\linewidth]{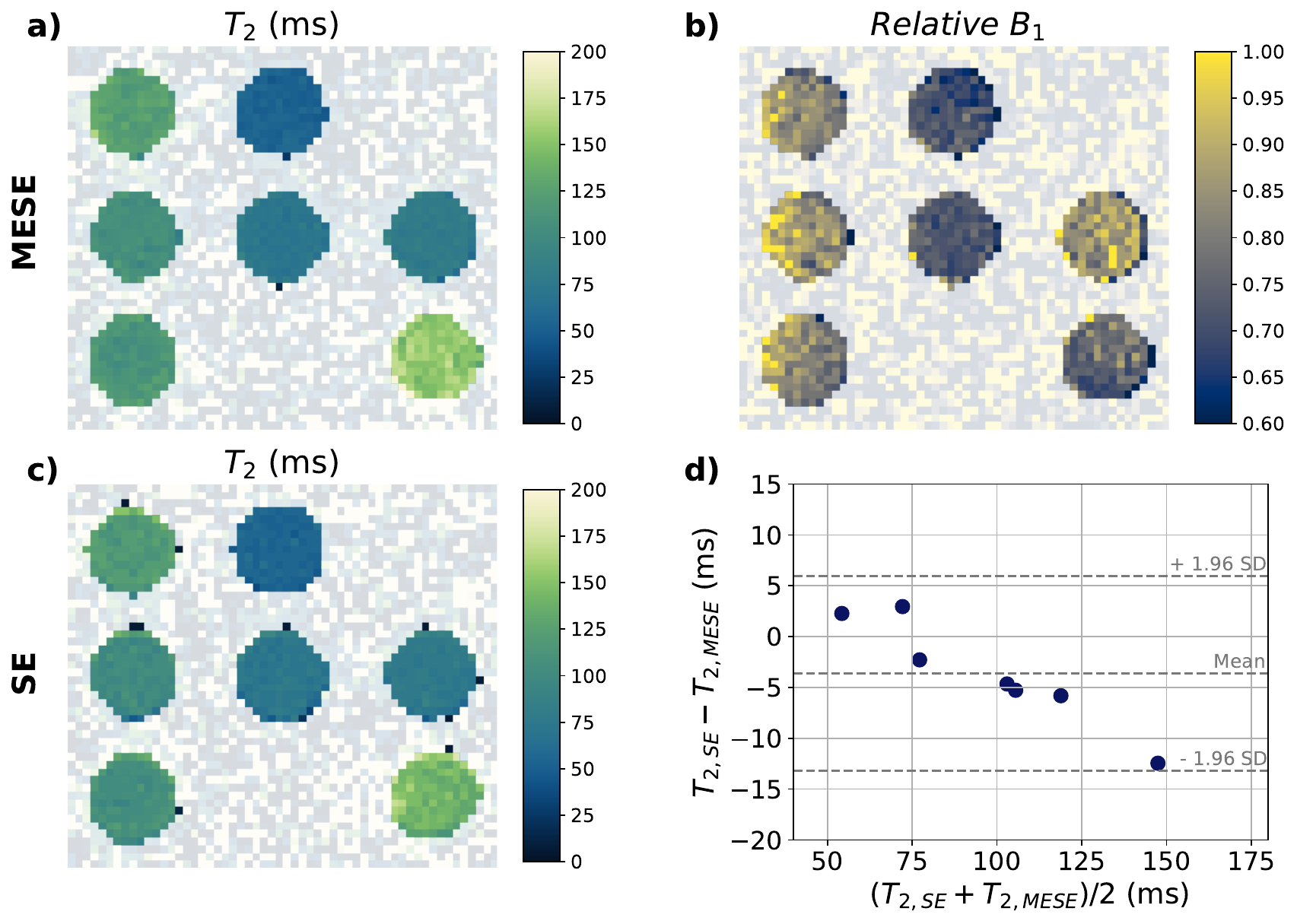}}
\caption{T$_{2}$-map (a) and relative B$_{1}$-map (b) from the MESE sequence, and T$_2$-map (c) from SE. The comparison (d) of T$_{2}$ values within each tube of the phantom demonstrates high agreement over the full range of values.}\label{fig:mrf}
\end{figure}

\autoref{fig:mrf}a and c show the T$_{2}$ maps of the phantom obtained using the multi-echo spin-echo (MESE) and spin-echo (SE) sequences, respectively. The relative B$_1$-map for the MESE sequence is shown in \autoref{fig:mrf}b. The Bland-Altman plot in \autoref{fig:mrf}d shows good agreement between the results obtained by the SE and the MESE methods. The absolute difference in T$_{2}$ values between SE and MESE was \qty{12(7)}{\milli\second} using a constant B$_1$ value and was reduced to \qty{5(3)}{\milli\second} using the more complex EPG signal model.

\subsection{Model-based DL}
\label{subsec:modl}

\begin{figure}
    \centering
\includegraphics[width=0.75\linewidth]{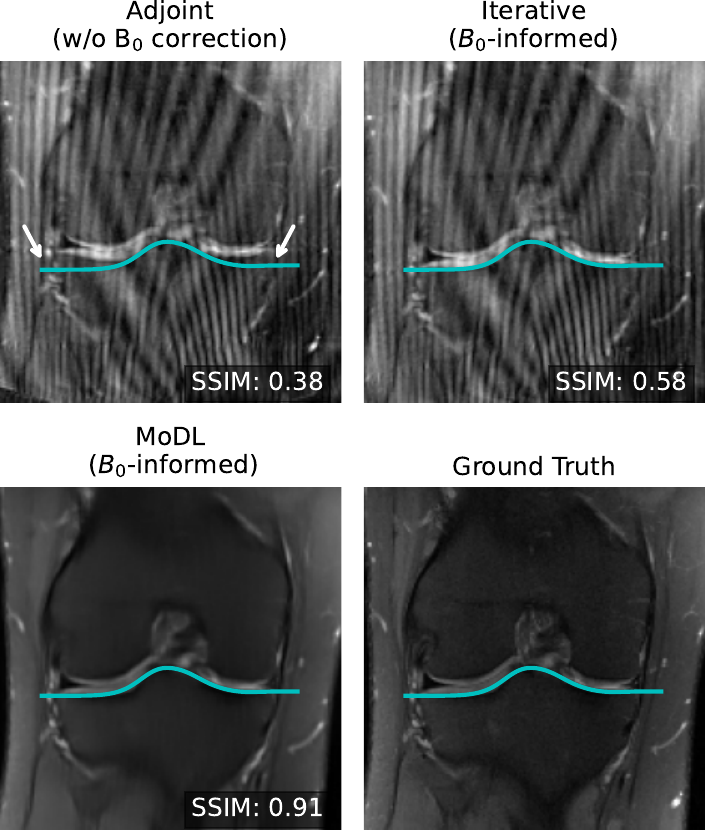}
    \caption{Comparison of reconstructed images from single-coil, 2-fold-accelerated fastMRI \cite{zbontar2018fastmri} knee data with simulated B$_0$-inhomogeneity using different methods. The cyan line marks the ground-truth tibia contour. The zero-filled Fourier transform without B$_0$ correction has strong artifacts and geometric distortion (arrows). The iterative reconstruction with B$_0$ correction removes the distortion. The B$_0$-informed MoDL \cite{aggarwal2018modl} reconstruction removes distortion and artifacts, resulting in the highest SSIM.}
    \label{fig:modl}
\end{figure}
\autoref{fig:modl} shows a B$_0$-informed MoDL \cite{aggarwal2018modl} reconstruction applied to a simulated 2-fold accelerated noisy acquisition with B$_0$ inhomogeneities and data based on the fastMRI knee dataset \cite{zbontar2018fastmri}. As baselines, we compare with a zero-filled adjoint reconstruction and a B$_0$-informed iterative reconstruction with Tikhonov regularization, all implemented in \texttt{MRpro}.
\rev{Across the test set ($N=50$), the B$_0$-informed MoDL achieved an SSIM of $0.90 \pm 0.05$ (mean $\pm$ standard deviation). In the example shown in \autoref{fig:modl}, the SSIM was 0.91, and MoDL reduced undersampling artifacts, geometric distortions, and intensity variations associated with B$_0$ inhomogeneity.}

\subsection{End-to-end training of physics-informed quantitative imaging}
\label{subsec:pinqi}

\begin{figure}[t]
\centering
\includegraphics[width=\linewidth,trim=0.0cm 0 0 0, clip]{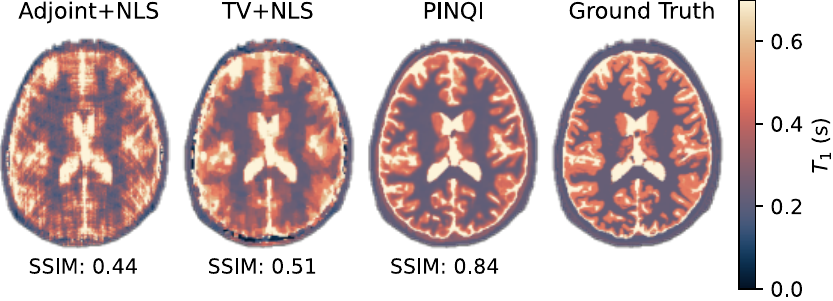}
\caption{Results for an implementation of \textit{PINQI} \cite{pinqi}: On a synthetic BrainWeb-based \cite{brainweb} saturation-recovery T$_1$-mapping dataset at 6-fold Cartesian undersampling, the physics-informed approach to learned quantitative imaging outperforms the classic approaches, i.e., adjoint or TV-regularized reconstruction followed by non-linear least squares.}\label{fig:pinqi}
\end{figure}

The results of T$_1$-mapping from a 6-fold undersampled saturation-recovery sequence using PINQI are shown in \autoref{fig:pinqi}. \rev{Across 24 test samples, PINQI achieved a T$_1$-map SSIM of $0.87 \pm 0.04$ and nRMSE of $0.045 \pm 0.005$ (mean $\pm$ standard deviation), compared with adjoint reconstruction followed by non-linear least squares (NLS; SSIM $0.39 \pm 0.04$, nRMSE $0.124 \pm 0.013$) and TV reconstruction followed by NLS (SSIM $0.46 \pm 0.06$, nRMSE $0.124 \pm 0.016$).} In the displayed example, adjoint reconstruction followed by NLS retained undersampling artifacts (SSIM $0.44$, nRMSE $0.19$), while TV reconstruction produced block artifacts and reduced resolution (SSIM $0.51$, nRMSE $0.12$). PINQI produced a high-fidelity T$_1$ map (SSIM $0.84$, nRMSE $0.04$) from the noisy, undersampled, single-coil data.

\subsection{\rev{Noise2Noise-trained reconstruction of
low-field data}}
\label{subsec:m4raw_results}

\begin{figure}[t]
    \centering
    \includegraphics[width=0.75\linewidth]{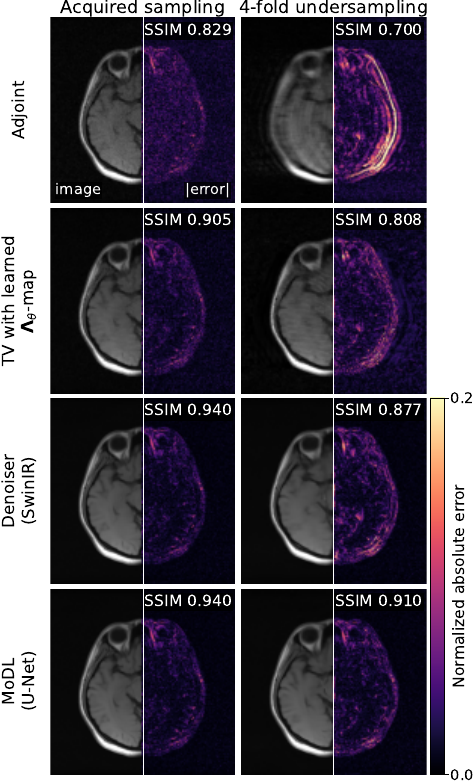}
    \caption{\rev{Repeat-supervised reconstruction of measured M4Raw data at the
    acquired sampling pattern (195 phase-encoding steps, left) and after retrospective four-fold
    undersampling (49 phase-encoding steps, right). Each tile combines the reconstruction (left) with
    its normalized absolute error (right). SSIM and errors use the registered
    average of repetitions 2--6 as reference.}}
    \label{fig:m4raw}
\end{figure}

\rev{\autoref{tab:reconstruction_metrics} reports metrics on the held-out test set for the acquired sampling pattern and retrospective four-fold Gaussian variable-density Cartesian undersampling.}
\rev{Among the evaluated methods, MoDL with a U-Net regularizer had the highest mean SSIM and PSNR at both the acquired sampling pattern (0.896 and 33.4 dB) and four-fold undersampling (0.857 and 30.6 dB); the SwinIR denoiser yielded similar values for acquired sampling (0.895 and 33.3 dB).}

\begin{table}[htpb]
\centering
\caption{\rev{Brain-masked test-set SSIM and PSNR (mean $\pm$ standard deviation across 25 subjects) for the M4Raw reconstructions.}}
\label{tab:reconstruction_metrics}
\setlength{\tabcolsep}{4pt}
\begin{tabular}{l c c c c}
\toprule
& \multicolumn{2}{c}{\textbf{Acquired Sampling}} & \multicolumn{2}{c}{\textbf{Four-Fold Undersampling}} \\
\cmidrule(lr){2-3} \cmidrule(lr){4-5}
\textbf{Method} & \textbf{SSIM} & \textbf{PSNR} & \textbf{SSIM} & \textbf{PSNR} \\
\midrule
Adjoint               & $0.703{\pm}0.022$ & $29.6{\pm}0.9$ & $0.619{\pm}0.015$ & $24.5{\pm}0.6$ \\
Scalar TV             & $0.736{\pm}0.020$ & $30.1{\pm}0.9$ & $0.643{\pm}0.017$ & $26.2{\pm}0.6$ \\
Adaptive TV & $0.821{\pm}0.016$ & $31.6{\pm}1.0$ & $0.722{\pm}0.016$ & $27.6{\pm}0.7$ \\
SwinIR                & $0.895{\pm}0.013$ & $33.3{\pm}1.3$ & $0.822{\pm}0.014$ & $29.1{\pm}0.8$ \\
MoDL                  & $0.896{\pm}0.013$ & $33.4{\pm}1.3$ & $0.857{\pm}0.015$ & $30.6{\pm}0.9$ \\

\bottomrule
\end{tabular}
\end{table}

\section{Discussion}

In this work, we introduced \texttt{MRpro}, a modular, open-source framework designed to bridge the gap between advanced MRI reconstruction and \rev{Python-based machine learning.}

Several software packages address aspects of MR reconstruction.
\rev{\texttt{MRpro} provides a unified, PyTorch-native environment with MR-specific data structures, composable operators, and tools for combining classical reconstruction algorithms with DL approaches.}
High-performance C/\CXX\ frameworks like BART \cite{bart} and Gadgetron \cite{gadetron} offer computational speed, but they present a significant barrier to extension, especially for ML researchers accustomed to Python. Conversely, general-purpose Python libraries for inverse problems, such as ODL \cite{odl} and DeepInverse \cite{deepinverse}, provide composable operators but lack the specialized, out-of-the-box support for MRI-specific data structures, operators (e.g., non-Cartesian trajectories, partial Fourier), and processing workflows. Other Python toolboxes are more closely related but have distinct goals. SigPy \cite{sigpy} provides excellent signal processing blocks but is not natively built on PyTorch, creating friction for deep learning integration. DIRECT \cite{direct} focuses on comparisons of DL methods and application to Cartesian reconstruction challenges, less on providing building blocks for general MR reconstruction tasks. MONAI \cite{cardoso2022monaiopensourceframeworkdeep} mainly focuses on image generation \cite{monaigenerative} and image segmentation, offering many implementations of common networks for these tasks. TorchIO \cite{torchio} provides datasets of medical images and image augmentations. Complex-valued data and raw k-space data with accompanying metadata are not supported. Finally, only a few solutions exist for qMRI, such as PyQMRI \cite{pyqmri} and qMRLab \cite{qmrlab}, neither of which integrates with DL frameworks. Thus, while there exist many related software solutions, \texttt{MRpro} is specifically designed to fill these gaps, offering a unified, PyTorch-native environment for both classical and ML reconstruction research.

\rev{ \autoref{fig:compare_frameworks} shows visually comparable reconstructions and close numerical agreement between \texttt{MRpro} and the other reconstruction packages for the selected iterative SENSE tasks. This comparison measures agreement with the \texttt{MRpro} reconstruction and does not establish absolute accuracy against a ground-truth image. The larger differences for the non-Cartesian reconstruction may reflect framework-specific choices for additional parameters, such as interpolation-kernel shape and size. This was also shown in a recent reproducibility study \cite{maierCGSENSERevisitedResults2021}. We did not compare more advanced reconstruction methods because directly comparable implementations were not available across all four packages.}

The framework's commitment to open standards like ISMRMRD, DICOM, and the vendor-agnostic sequence standard Pulseq promotes reproducibility \cite{Tamir2025} and enables seamless integration with open-source research systems, such as the OSI$^2$ low-field MRI using the Nexus console \cite{schote2025} (\autoref{subsec:tv_maps}) while being fully compatible with proprietary clinical scanners. For Siemens systems, MRpro also offers an OpenRecon container that enables the usage of \texttt{MRpro} directly at the scanner.
\texttt{MRpro} is designed in a modular way starting with basic building blocks such as data objects, operators including signal models and solvers. These are combined into ready-to-use reconstruction pipelines (e.g., TV-regularized iterative image reconstruction or iterative SENSE). We have shown how custom algorithms can be created using the provided building blocks.

We demonstrated this for three diverse DL approaches: MoDL \cite{aggarwal2018modl} (\autoref{subsec:modl}), learned regularization maps \cite{kofler2023learning} (\autoref{subsec:tv_maps}), and PINQI \cite{pinqi} (\autoref{subsec:pinqi}). \rev{These examples implement established methods within a common differentiable framework.
We further evaluated these components on measured four-channel M4Raw data at 0.3\,T using repeat-supervised training without noise-free targets (\autoref{subsec:m4raw_results}).}

The effectiveness of the predefined neural networks was also validated by competitive performance \cite{zimmermann2025augmentaugmentdiverseaugmentations} in the 2025 ultra-low-field MICCAI challenge, which compared different approaches for denoising, super-resolution, and contrast translation from 3D multi-contrast LF images to co-registered high-field images.

For the MESE sequence used in \autoref{subsec:mrfmethod}, an accurate \qty{180}{\degree} refocusing pulse during the multi-echo train leads to a mono-exponential decay of the signal. Nevertheless, for deviations from this perfect refocusing pulse, the signal behavior is much more complex. The advanced EPG model can account for these system imperfections and ensure accurate T$_{2}$ estimation. This was shown by a reduced difference between the MESE and SE sequence of more than \qty{50}{\percent} from \qty{12(7)}{\milli\second} to \qty{5(3)}{\milli\second}.

The framework's utility for quantitative low-field MRI was \rev{demonstrated} through a dictionary-based ultra-low-field application (\autoref{subsec:mrf}) and noise-prone diffusion imaging (\autoref{subsec:diffusion}). In the latter, regularization reduced noise levels, yielding ADC values that more closely matched reference values. Remaining discrepancies may be attributed to the high noise floor of the acquisition and the unaccounted temperature dependence of the ADC values.

Previously, the framework secured a second-place award in the MRI Study Group challenge at ISMRM 2024 on T$_1$ and T$_{2}^*$ mapping. In that challenge, the data consisted of turbo spin-echo images obtained from a phantom at different time points after an inversion pulse \cite{tatman202410868350} and gradient-echo images with different echo times for T$_{2}^*$ mapping \cite{tatman202410868361}. Our entry used dictionary matching to obtain initial estimates of the relaxation times, followed by non-linear fitting of the signal models using Adam \cite{kingmaAdamMethodStochastic2015a}. Beyond relaxometry, \texttt{MRpro} recently powered an end-to-end trainable, physics-informed neural network for Magnetic Resonance Elastography \cite{martin_mre}.

\rev{While \texttt{MRpro} provides differentiable EPG and Bloch--McConnell models as composable operators for signal evaluation in quantitative MRI, MRzero \cite{Loktyushin2021} uses phase distribution graphs to simulate complete MR acquisitions from Pulseq sequence descriptions, including imaging gradients, and supports differentiable sequence optimization \cite{Endres2024}. The two frameworks were combined in a previous reproducibility study \cite{Tamir2025}, in which MRzero simulated the acquisition of a virtual phantom from a Pulseq sequence and \texttt{MRpro} reconstructed the resulting k-space data.}

\texttt{MRpro} has several limitations. \rev{\texttt{MRpro} requires Python and PyTorch runtimes. Deploying Python environments on vendor scanners was historically difficult. Yet on supported Siemens systems, \texttt{MRpro} can run through an OpenRecon container and the use of Python is already established in open-source low-field systems \cite{schote2025,NEGNEVITSKY2023107424}. Hardware compatibility is limited to platforms supported by
  PyTorch, including common CPU systems and several GPU backends. Supporting TensorFlow \cite{tensorflow} or JAX \cite{jax} would require a separate backend implementation.}
\rev{\texttt{MRpro}'s operator interface prioritizes composability, vectorization, and a simple interface, which can require more memory than problem-specific implementations, particularly for large 3D or dynamic datasets. We have not systematically benchmarked scaling for these workloads. The \texttt{LinearOperatorMatrix} can reduce peak memory use by evaluating smaller subproblems sequentially, at the cost of longer runtime.}
\rev{\texttt{MRpro} keeps k-space data aligned with its trajectory and metadata during all processing steps, but this does not establish that the imported acquisition information is correct. Incorrect headers or trajectories can configure the wrong acquisition model and must be checked by the user.}

The goal of this work is to present a reproducible and extensible reconstruction framework and representative low-field use cases, rather than to claim state-of-the-art performance on each individual task.

While \texttt{MRpro} provides a robust foundation, including essential components for pre-whitening, coil compression, and sensitivity estimation, it does not yet include all established methods for high-field data, such as ESPIRiT \cite{uecker2013} or GRAPPA \cite{Griswold2002}. The modular architecture and comprehensive continuous integration pipeline are explicitly designed to facilitate community contributions. We therefore invite the research community to extend \texttt{MRpro} by integrating these and other advanced methods, further enhancing its value as a shared tool for open and reproducible science.

\section{Conclusions}

We presented an open-source image reconstruction package that uses PyTorch to allow for fast processing and native integration with DL approaches. \texttt{MRpro} follows a highly modular structure with low-level operators and high-level image reconstruction algorithms. We demonstrated the capabilities of \texttt{MRpro} on a range of different reconstruction problems including model-based DL with correction of B$_0$-inhomogeneities and end-to-end quantitative imaging. All examples can be reproduced by freely available Jupyter notebooks and open-access data on Zenodo. The comprehensive quality control and testing framework of \texttt{MRpro} follows modern software development standards and \rev{supports maintainability and long-term development.}

\bmsection*{Author contributions}
FFZ, PS and CK led the project. FFZ designed and implemented the software, carried out experiments and visualizations, and supervised software development. PS and CK supervised the work, contributed to software design and implementation, and supported data acquisition. CSA, HH, CRK, JS, DS, and NJ acquired data. AK, CRK, and HH carried out experiments and visualizations. DS and LL contributed visualizations. AK, CRK, MG, SM, LL, BAB, DS, and JH contributed to the software. All authors discussed the results, reviewed and approved the manuscript.

\bmsection*{Acknowledgments}
We would like to thank Tom O'Reilly and Andrew Webb for providing the in vivo data acquired on the OSI$^2$ ONE open-source low-field scanner.

\bmsection*{Conflict of interest}

The authors declare no potential conflicts of interest.

\bmsection*{Data Availability Statement}
\texttt{MRpro} is open-source and available at \url{https://github.com/PTB-MR/mrpro}. \rev{Upon acceptance for publication, we will tag the exact software version used for this work as \texttt{nmr\_in\_biomed}.}
The package is distributed via PyPI. The data required to reproduce the results are freely available via Zenodo. We provide Jupyter notebooks demonstrating the functionality in more detail\footnote{\url{https://docs.mrpro.rocks/examples.html}}. Each notebook can be run in Google Colab directly from the browser. \rev{The notebooks are structured as tutorials to explain how to use \texttt{MRpro} in a step-by-step way and provide the mathematical background of the different reconstruction algorithms. High-level functions such as regularized iterative SENSE or TV-regularized image reconstruction are also decomposed into their low-level building blocks to give users more detailed insight.}
\bibliography{references}

\end{document}